\documentclass[12pt]{article}

\usepackage{t1enc}
\usepackage[latin1]{inputenc}
\usepackage[english]{babel}

\usepackage{graphicx}

\begin{document}


\title{
Model and simulation of wide-band interaction
in free-electron lasers
        }

\author{Yosef Pinhasi
\footnote{\em E-mail adress: yosip@eng.tau.ac.il}
, Yuri Lurie, and Asher Yahalom\\[3mm]
The College of Judea and Samaria\\[1mm]
Dept. of Electrical and Electronic Engineering\\
--- Faculty of Engineering\\[1mm]
P.O. Box 3, Ariel 44837, Israel
          }


\maketitle

\begin{abstract}

A three-dimensional, space-frequency model for simulation of interaction in free-electron lasers (FELs)
is presented. The model utilizes an expansion of the total electromagnetic field
(radiation and space-charge waves) in terms of transverse eigenmodes of the waveguide,
in which the field is excited and propagates.
The mutual interaction between the electron beam and the electromagnetic field is fully described
by coupled equations, expressing the evolution of mode amplitudes and electron
beam dynamics.

Based on the three-dimensional model, a numerical particle simulation code was developed.
A set of coupled-mode excitation equations, expressed in the frequency domain,
are solved self-consistently with the equations of particles motion.
Variational numerical methods were used to simulate excitation of backward modes.
At present, the code can simulate free-electron lasers operation in various modes:
spontaneous (shot-noise) and self-amplified spontaneous emission (SASE),
super-radiance and stimulated emission, all in the non-linear Compton or Raman regimes.

\end{abstract}


\section{Introduction}

Several numerical models have been suggested for three-dimensional simulation
of the FEL operation in the non-linear regime
\cite{cite1}-\cite{cite10}.
Unlike a previously developed steady-state models, in which the interaction is assumed
to be at a single frequency (or at discrete frequencies), the approach presented in this paper
considers a continuum of frequencies, enabling solution of non-stationary, wide-band interactions
in electron devices operating in the linear (small-signal) and non-linear (saturation) regimes.
Solution of excitation equations in the space-frequency domain inherently takes
into account dispersive effects arising from cavity and beam loading.
The model is based on a coupled-mode approach expressed in the frequency domain \cite{Pinhasi_Gover_PhysRevE93}
and used in the WB3D particle simulation code to calculate the total
electromagnetic field excited by an electron beam drifting along a waveguide in
the presence of a wiggler field of FEL.

\begin{figure}[t]
\centerline{
\includegraphics[width=0.7\textwidth,angle=0]{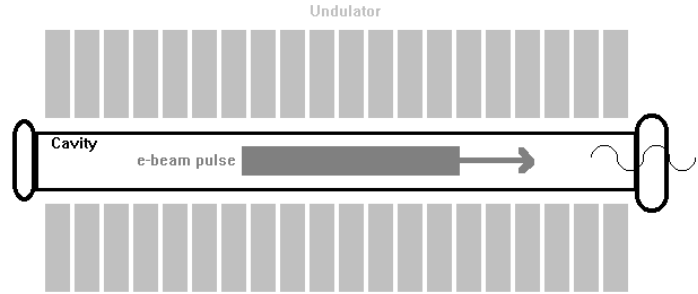}
                   }

\vspace{-10mm}
\caption{The FEL scheme}
\label{scheme}
\end{figure}

The unique features of the present model enable one to solve non-stationary
interactions taking place in electron devices such as spontaneous and
super-radiant emissions in a pulsed beam FEL, shown in Fig.\ \ref{scheme}.
We employed the code to demonstrate a spontaneous and super-radiant emissions excited
when a bunch of electrons passes through a wiggler of an FEL.
Calculations of the power and energy spectral distribution in the frequency domain were carried out.
The temporal field was found by utilizing a procedure of inverse Fourier transformation.
Super-radiance in the special limit of 'grazing'
(where dispersive waveguide effects play a role) was also investigated.

\section{Dynamics of the particles}

The state of the particle $i$ is described by a six-components vector,
which consists of the particle's position coordinates
${\bf r}_i = (x_i,y_i,z_i)$ and velocity vector ${\bf v}_i$.
Here $(x,y)$ are the transverse coordinates and $z$ is the
longitudinal axis of propagation.  The velocity of each particle, in
the presence of electric ${\bf E}({\bf r},t)$
and magnetic ${\bf B}({\bf r},t) = \mu {\bf H}({\bf r},t)$  fields,
is found from the Lorentz force equation:
\begin{equation}
\frac{d}{d t} \left( \gamma_i  {\bf v}_i  \right)
= - \, \frac{e}{m}
\ \left[
{\bf E}({\bf r}_i,t) + {\bf v}_i \times {\bf B}({\bf r}_i,t)
\right]
\end{equation}
where
$e$ and $m$ are the electron charge and mass respectively.
The fields represent the total (DC and AC) forces operating on the
particle, and include also the self-field due to space-charge.
The Lorentz relativistic factor $\gamma_i$ of each particle is found
from the equation for kinetic energy:
\begin{equation}
\frac{d \gamma_i}{d t} = - \, \frac{e}{m c^2} \  {\bf v}_i \cdot {\bf E}({\bf r}_i,t)
\end{equation}
where $c$ is the velocity of light.

The equations are rewritten, such that the coordinate of the
propagation axis $z$ becomes the independent variable, by replacing
the time derivative $\frac{d}{d t}=v_{z_i} \frac{d}{d z}$.
This defines a transformation of variables for each particle, which
enables one to write the three-dimensional equations of motion in
terms of $z$:
\begin{equation}
\frac{d {\bf v}_i}{d z} =
\frac{1}{\gamma_i}
\left\{- \, \frac{e}{m} \, \frac{1}{v_{z_i}}
\left[ {\bf E}({\bf r}_i,t) + {\bf v}_i \times {\bf B}({\bf r}_i,t) \right]
  -  {\bf v}_i \, \frac{d \gamma_i}{d z} \right\}
                       \label{force_velocity}
\end{equation}

\begin{equation}
\frac{d\gamma_i}{d z} =
- \, \frac{e}{m c^2} \, \frac{1}{v_{z_i}} \ {\bf v}_i \cdot {\bf E}
                                                     \label{force_gamma}
\end{equation}

The time it takes a particle to arrive at a position $z$, is
a function of the time $t_{0i}$ when the particle entered at $z=0$, and
its instantaneous longitudinal velocity $v_{zi}(z)$ along the path of motion:
\begin{equation}
t_i(z)=t_{0_i}+\int_0^z \frac{1}{v_{z_i}(z')}d z'
\label{time}
\end{equation}

\section{The driving current}

The distribution of the current in the beam is determined by the
position and the velocity of the particles:
\begin{eqnarray}
{\bf J} ({\bf r},t)
&=& - \sum\limits_{i}
  q_i \, {\bf v}_i \, \delta (x-x_i) \, \delta (y-y_i) \ \delta \left[z - z_i(t)  \right] =
                              \nonumber \\
&=& - \sum\limits_{i}
  q_i \, \left(\frac{{\bf v}_i}{v_{z_i}}\right) \, \delta (x-x_i) \ \delta (y-y_i) \,
  \delta \left[t - t_i(z)\right]
\end{eqnarray}
here
$q_i$ is the charge of the $i$th macro particle in the simulation.
The Fourier transform (in the positive frequency domain)
of the current density is given by:
\begin{eqnarray}
{\widetilde{\bf J}}({\bf r},f) &=&
2 u(f) \int\limits_{-\infty}\limits^{+\infty}
       {\bf J}({\bf r},t)  e^{-j 2 \pi f t}  dt  =   \nonumber \\
 &=&
- 2  u(f)  \sum\limits_{i}
q_i \left( \frac{{\bf v}_i}{v_{z_i}} \right)
          \, \delta (x-x_i) \, \delta (y-y_i) \ e^{-j 2\pi  f t_i(z)}
\label{current_Fourier}
\end{eqnarray}
here
$
u(f) = \left\{
          \begin{array}{ll}
               1 ,  &  f \ge 0 \\
               0 ,  &  f < 0
          \end{array}\right.
$
is the step function.

This Fourier transform of the current (\ref{current_Fourier})
is substituted in the following excitation equations
to find the evolution of the electromagnetic fields.

\section{The electromagnetic field}

The Fourier transform of the transverse component of the total electromagnetic
field is given at the frequency domain
as a superposition of waveguide transverse eigenmodes

\begin{eqnarray}
{\widetilde{\bf E}} _{\perp} ({\bf r},f)
&=& \sum\limits_q \left\{
     C_{+q}(z,f)  e^{-j k_{z_q} z}
      +  C_{-q}(z,f)   e^{+j k_{z_q} z} \right\}
                {\widetilde{\mathcal E}}_{ q _{\perp}}(x,y)  \nonumber  \\
\widetilde{\bf H} _{\perp} ({\bf r},f)
&=& \sum\limits_q \left\{
      C_{+q}(z,f)  e^{-j k_{z_q} z}
   -  C_{-q}(z,f) e^{+j k_{z_q} z} \right\}
                  {\widetilde{\mathcal H}}_{q _{\perp}}(x,y)
\label{EMfield_transverse}
\end{eqnarray}
and the expression for the longitudinal component of the
electromagnetic field is found to be:
\begin{eqnarray}
\widetilde{E} _{z} ({\bf r},f)
&=& \sum\limits_q  \left\{
      C_{+q}(z,f)  e^{-j k_{z_q} z}
   -  C_{-q}(z,f) e^{+j k_{z_q} z} \right\}
       \widetilde{\mathcal E}_{ q _z}  (x,y)
       +  \frac{j}{2\pi f \varepsilon}
                  \widetilde{J}_z ({\bf r},f)  \nonumber  \\
\widetilde{H} _z ({\bf r},f)
&=& \sum\limits_q  \left\{
      C_{+q}(z,f)  e^{-j k_{z_q} z}
   +  C_{-q}(z,f) e^{+j k_{z_q} z} \right\}
                {\widetilde{\mathcal H}}_{q _z}  (x,y)
\label{EMfield_z}
\end{eqnarray}
Where
$k_{z_q} = \sqrt{ \left( \frac{2\pi f}{c} \right)^2  -  k_{\perp _q}^2 }$
($k_{\perp _q}$ is the cut-off wave number of mode $q$)
and
$C_{+q}(z,f)$ and $C_{-q}(z,f)$ are the $q$th mode's amplitude
corresponding to the forward and backward waves, respectively.
Equations (\ref{EMfield_transverse}) and (\ref{EMfield_z}) describe the total
transverse and longitudinal electromagnetic field (radiation and space-charge
waves) \cite{Pinhasi_Gover_PhysRevE93}.

The evolution of the $q$th mode amplitudes $C_{\pm q}(z,f)$ is found
after substitution of the current distribution
(\ref{current_Fourier}) into the scalar differential excitation equation:
\begin{eqnarray}
\lefteqn{
\frac{d}{d z} C_{\pm q}(z,f) =}  \nonumber \\
&=&  \mp \frac{1}{2 {\mathcal N}_{q}(f)}  e^{\pm j  k_{z_q}  z}
    \int \int       \left[
 \left(\frac{Z_q}{Z_q ^\ast}\right) {\widetilde{\bf J}}_{\perp}({\bf r},f) \ +
     \hat{{\bf z}} \widetilde{J}_{z}({\bf r},f)
       \right] \cdot { {\widetilde{\mathcal E}}_{\pm q} }^\ast (x,y) \,  d x \, d y  =  \nonumber  \\
&=&
\pm  \frac{1}{{\mathcal N}_q(f)}  e^{\pm j  k_{z_q}  z}
 \sum\limits_{i}    q_i  \;  e^{-j  2\pi  f  t_i(z)}
\left\{
   \frac{\zeta_q}{v_{z_i}}  {\bf v}_{\perp _i}  \cdot
             { \widetilde{\mathcal E}_{\pm q _\perp} }^\ast (x_i,y_i)
   +        { \widetilde{\mathcal E}_{\pm q _z} }^\ast (x_i,y_i)   \right\}
                                                                               \nonumber  \\
                              \label{excitation_Coefficients}
\end{eqnarray}
here
\[
{\mathcal N}_q(f) =
\int \int
\left[
\widetilde{{\mathcal E}}_{q \perp}
\times
\widetilde{{\mathcal H}}_{q \perp}^\ast \right]
\cdot \hat{\bf z}     \     d x \, d y
\]
is the power normalization of mode $q$, and
\[
\zeta _q \equiv  \frac{Z_q}{Z_q ^\ast}
  = \left\{ \begin{array}{ll}
                   +1, & propagating\ modes \\
                    -1,  & cut-off\ modes
              \end{array}\right.
\]

The total electromagnetic field is found by inverse Fourier transformation of
(\ref{EMfield_transverse}) and (\ref{EMfield_z}):
\begin{eqnarray}
{\bf E} ({\bf r},t) &=&
\Re \left\{  \int\limits_{0}\limits^{\infty}
{\widetilde{\bf E}}({\bf r},f) \ e^{+j 2\pi f t} \, d f \right\}  \nonumber  \\
{\bf H}({\bf r},t) &=&
\Re \left\{  \int\limits_{0}\limits^{\infty}
{\widetilde{\bf H}}({\bf r},f) \  e^{+j 2\pi f t} \, d f \right\}
\end{eqnarray}

The energy flux spectral distribution (defined in the positive frequency domain $f \ge 0$)
is given by:
\begin{eqnarray}
\frac{d W(z)}{d f} &=&
\frac{1}{2}   \Re\left\{  \int\int  \left[ {\widetilde{\bf E}}({\bf r},f)
          \times   {\widetilde{\bf H}}^{\ast}({\bf r},f) \right]
                \cdot  \hat{{\bf z}} \ d x \, d y  \right\}   =    \nonumber  \\
&=&
\frac{1}{2}
\sum_q^{Propagating} \left[ |C_{+q}(z,f)|^2 - |C_{-q}(z,f)|^2 \right] \Re\left\{ {\mathcal N}_q(f) \right\}
                                                                \nonumber  \\
&& \hspace{2cm}
+
\sum_q^{Cut-off}
\Im \left\{ C_{+q}(z,f)  C_{-q}^\ast(z,f) \right\}   \Im \left\{ {\mathcal N}_q(f) \right\}
\end{eqnarray}

\section{The Variational Principle}

The solution of the equations
(\ref{force_velocity}), (\ref{force_gamma}) and (\ref{excitation_Coefficients})
for forward waves is done by integrating the equations in the positive
$z$-direction for a given boundary conditions at the point $z=0$.
For backward waves the natural physical boundary conditions are given
at the end of the interaction region $z=L_w$
and the direction of the integration is the negative $z$-direction.

In order to take into account excitation of both forward and backward waves,
we introduce a variational functional
\begin{eqnarray}
\lefteqn{F =}&&
                           \nonumber \\
&=&\int_0^{L_w} \left[ C_{-q}(z,f) \, \frac{dC_{+q}(z,f)}{d z}
\ -\  C_{+q}(z,f) \, \alpha _q(z,f)
     \ +\  C_{-q}(z,f) \, \beta _q(z,f) \right] d z
                           \nonumber \\
\label{F-var}
\end{eqnarray}
where
\begin{eqnarray}
\lefteqn{\alpha _q (z,f) =} && \nonumber \\
&&  -\frac{1}{2 {\mathcal N}_{q}(f)}  e^{- j  k_{z_q}  z}
    \int \int       \left[
 \left(\frac{Z_q}{Z_q ^\ast}\right) {\widetilde{\bf J}}_{\perp}({\bf r},f) \ +
     \hat{{\bf z}} \widetilde{J}_{z}({\bf r},f)
       \right] \cdot { {\widetilde{\mathcal E}}_{-q} }^\ast (x,y) \,  d x \, d y
\nonumber \\[7mm]
\lefteqn{\beta _q (z,f) =} && \nonumber \\
&& + \frac{1}{2 {\mathcal N}_{q}(f)}  e^{+j  k_{z_q}  z}
    \int \int       \left[
 \left(\frac{Z_q}{Z_q ^\ast}\right) {\widetilde{\bf J}}_{\perp}({\bf r},f) \ +
     \hat{{\bf z}} \widetilde{J}_{z}({\bf r},f)
       \right] \cdot { {\widetilde{\mathcal E}}_{+q} }^\ast (x,y) \,  d x \, d y
\nonumber \\
\label{alpha-beta}
\end{eqnarray}

The variational derivative of the above functional is:
\begin{eqnarray}
\lefteqn{
\delta F = \int_0^{L_w} \left[ \delta C_{-q}(z,f) \, \left( \frac{dC_{+q}(z,f)}{d z} + \beta _q(z,f) \right)
\right.     } &&                              \nonumber \\
&&  
\left.
\ -\  \delta C_{+q}(z,f) \, \left( \frac{dC_{-q}(z,f)}{d z} + \alpha _q(z,f) \right)
\ + \ \frac{d (C_{-q}(z,f) \, \delta C_{+q}(z,f))}{d z} \right] d z          \nonumber \\
\label{F-variated}
\end{eqnarray}

For arbitrary variations $\delta C_{\pm q}(z,f)$ the functional minimizes
(i.e., $\delta F = 0$) if and only if the equations
(\ref{excitation_Coefficients}) are satisfied, and the boundary term is
\begin{equation}
\delta F_B =
\int_0^{L_w} \frac{d \left( C_{-q}(z,f) \, \delta C_{+q}(z,f) \right)}{d z} d z
= \left.  C_{-q}(z,f) \ \delta C_{+q}(z,f) \right| _0^{L_w} = 0
\label{delta FB}
\end{equation}
resulting in
\begin{equation}
C_{-q}(0,f) \, \delta C_{+q}(0,f) = C_{-q}(L_w,f) \, \delta C_{+q}(L_w,f)
\label{bounc}
\end{equation}

This enable solving amplifier scheme in which the boundary conditions are
$C_{-q}(L_w,f)=0$ and $C_{+q}(0,f)=0$, as well as
an oscillator configuration where the boundary conditions are
$C_{-q}(0,f) = C_{-q}(L_w,f)$ and $C_{+q}(0,f)=C_{+q}(L_w,f)$.

\section{Numerical results}

We shall use the code to investigate super-radiant emission radiated when an
ultra short $e$-beam bunch (with duration of 1~pS, much shorter than the temporal period
of the signal) passes through the wiggler of an FEL having operational
parameters as given in table\ \ref{table}.
In this case, the power of super-radiant (coherent) emission is much higher
than that of the incoherent spontaneous emission \cite{cite11}.

\begin{figure}[b]
\centerline{
\includegraphics[width=0.7\textwidth,angle=0]{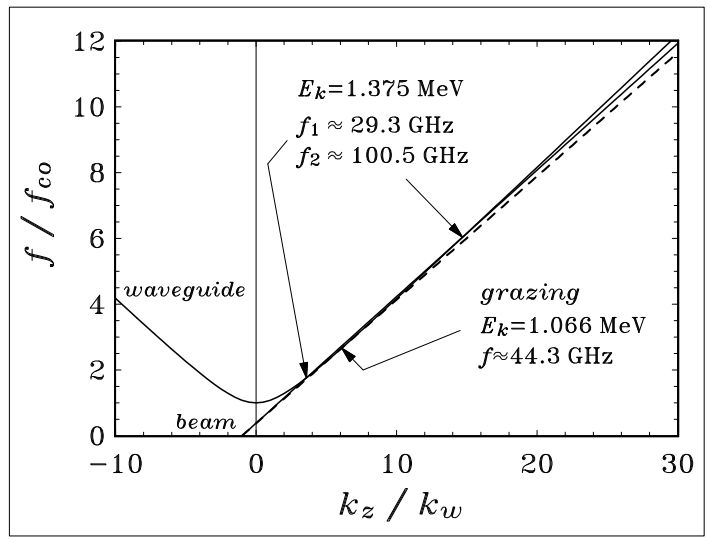}
                   }

\caption{FEL dispersion curves}
\label{dispersion}
\end{figure}

Fig.\ \ref{dispersion} shows two cases of dispersion relations: when the beam
energy is set to 1.375~MeV, there are two separated intersection points between
the beam and waveguide dispersion curves, corresponding to the ``slow'' ($v_{g_1} < v_{z_0}$)
and ``fast'' ($v_{g_2} > v_{z_0}$) synchronism frequencies 29~GHz and 100~GHz,
respectively. Lowering the beam energy to 1.066~MeV, results in a single
intersection (at 44~GHz), where the beam dispersion line is tangential to the
waveguide dispersion curve ($v_{g} = v_{z_0}$ --- ``grazing limit'').

\begin{figure}[t]
\centerline{
\includegraphics[width=0.48\textwidth,angle=0]{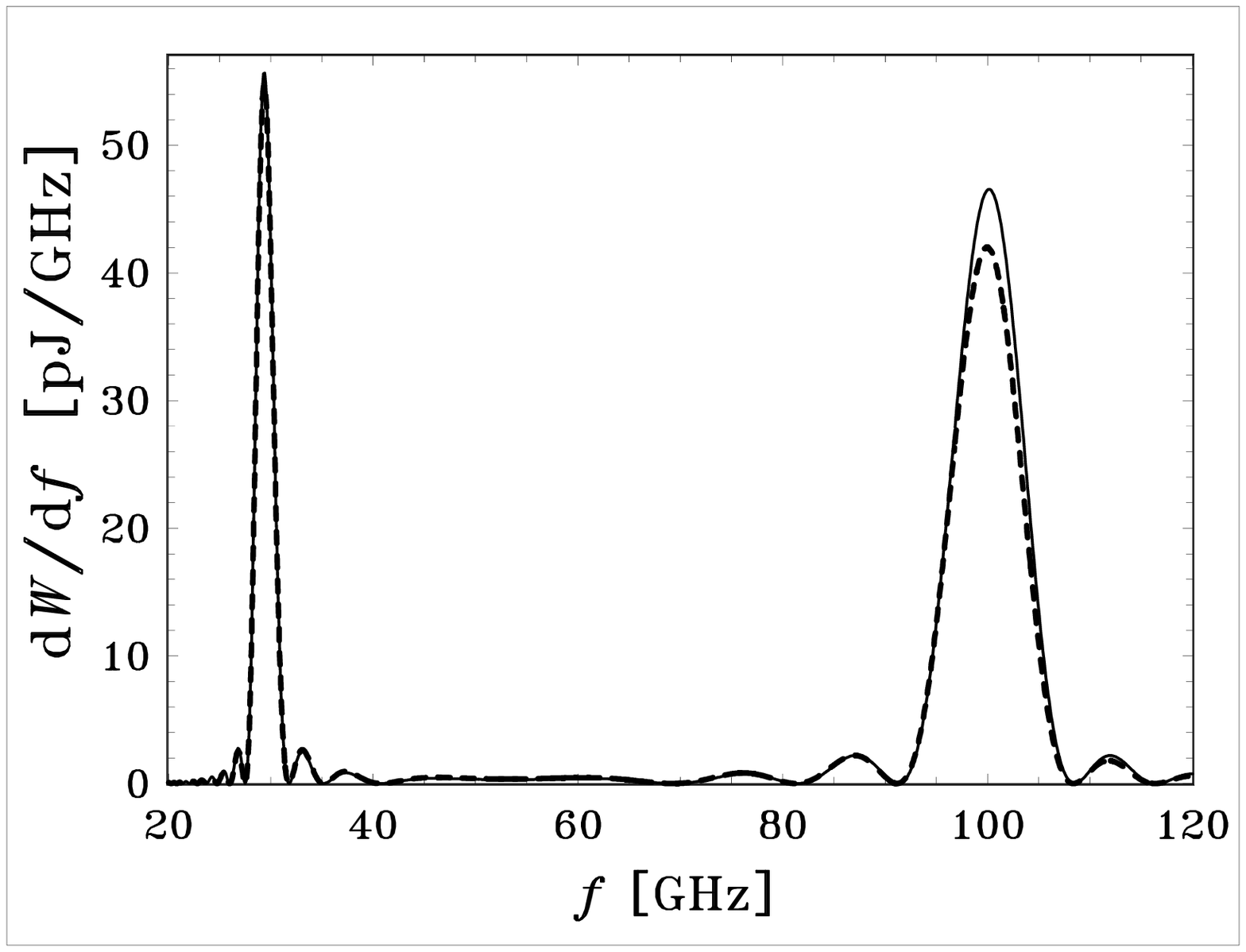}
\hfill
\includegraphics[width=0.48\textwidth,angle=0]{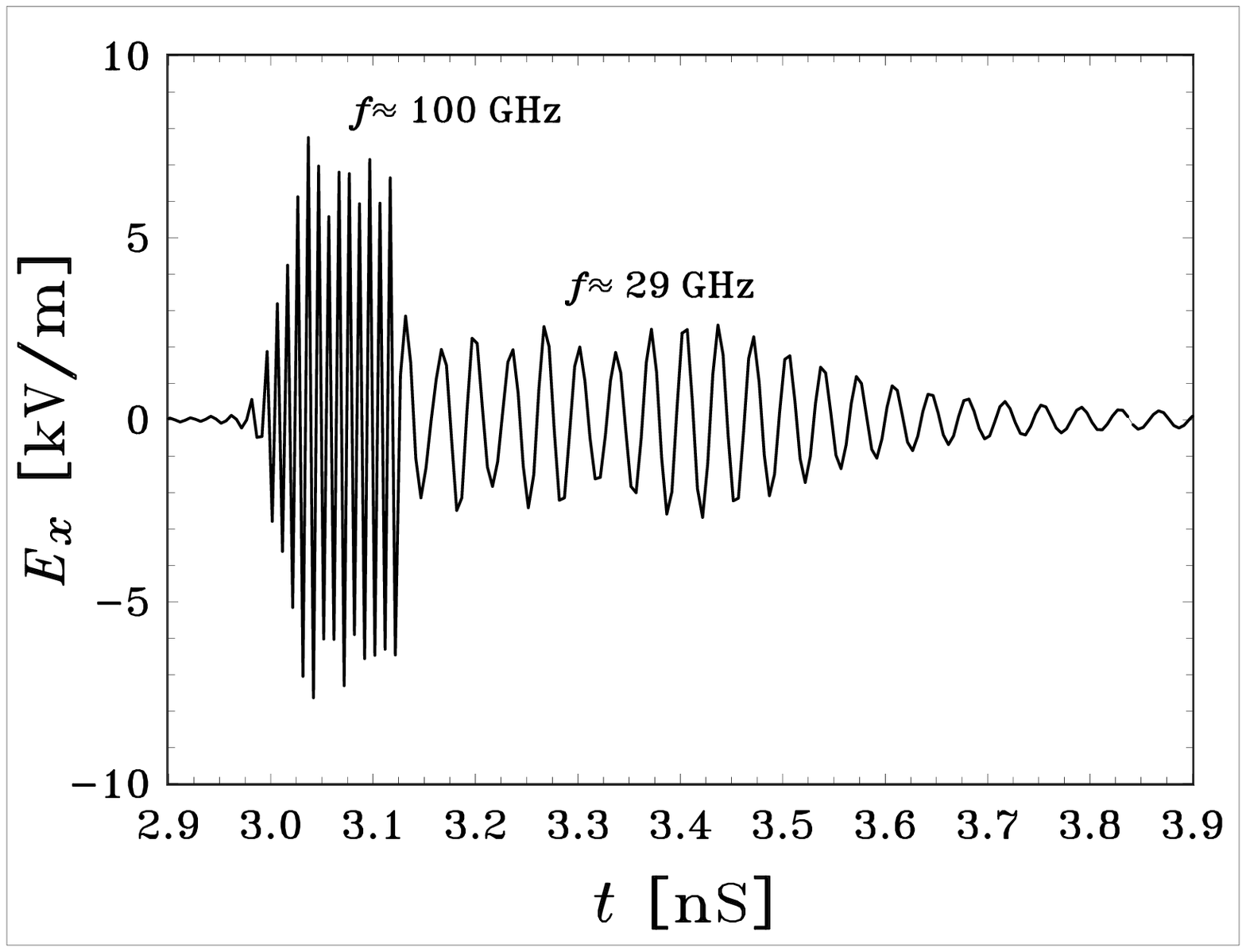}
                   }

\caption{
Super-radiant emission from an ultra short bunch:
(a)~Energy spectrum (analytic calculation and numerical simulation are shown by solid and dashed lines,
respectively);
(b)~temporal wavepacket.
         }
\label{two_solutions_energy}
\end{figure}

\begin{figure}[b]
\centerline{
\includegraphics[width=0.48\textwidth,angle=0]{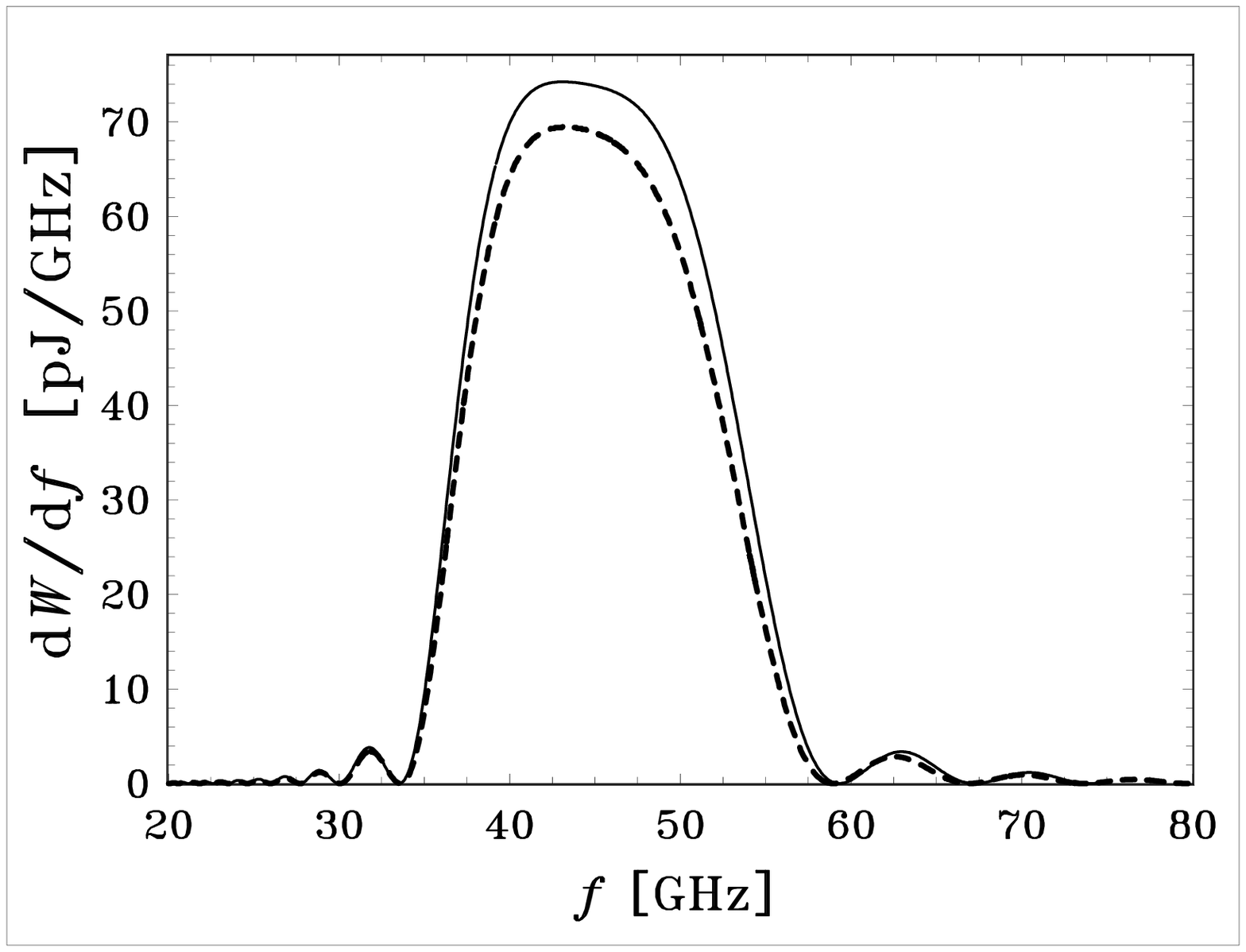}
\hfill
\includegraphics[width=0.48\textwidth,angle=0]{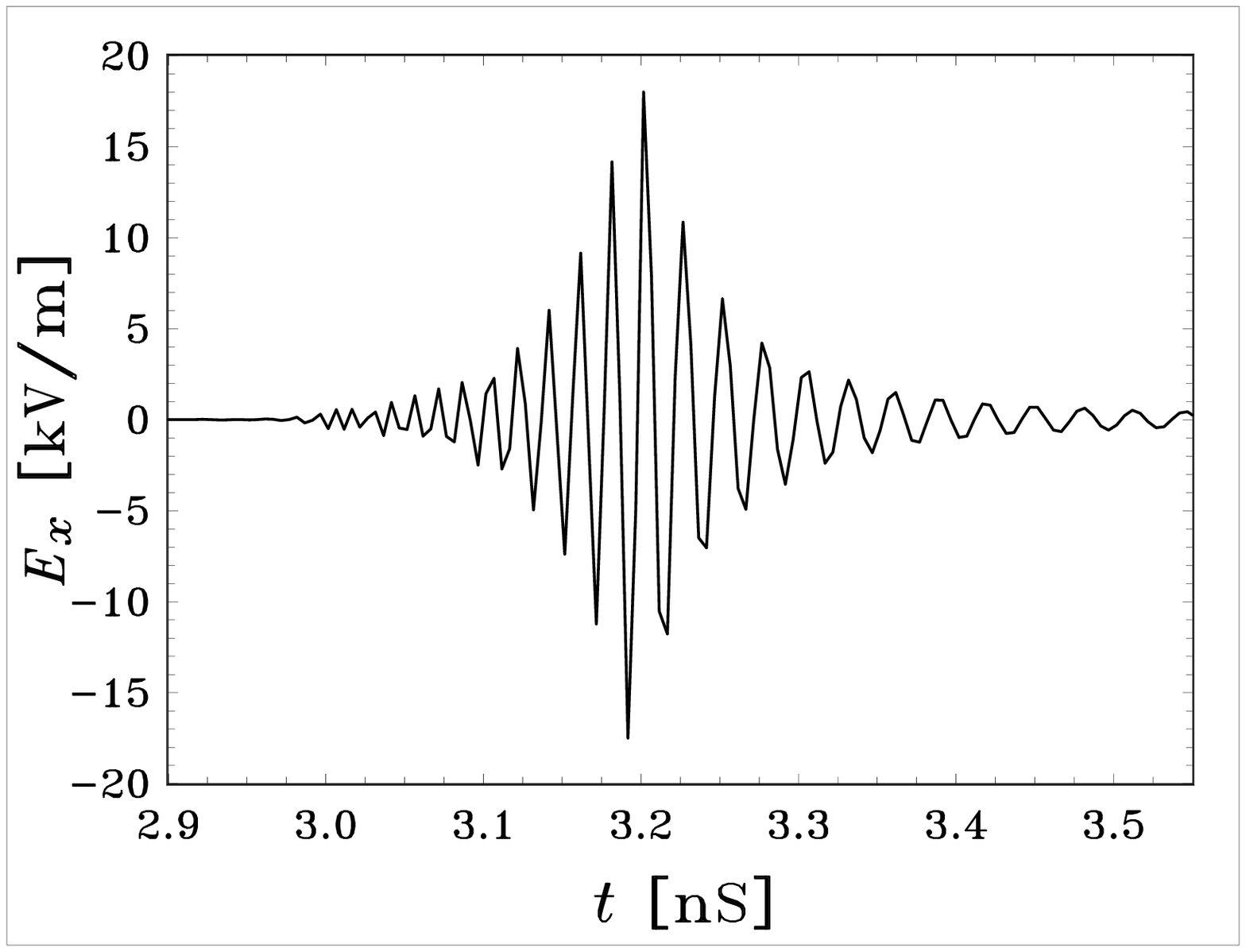}
                   }

\caption{
That of Fig.\ \ref{two_solutions_energy}, but in the grazing limit.
         }
\label{grazing_energy}
\end{figure}

The calculated spectral density of energy flux in the case of two
well-separated solutions is shown in Fig.\ \ref{two_solutions_energy}a.
The spectrum peaks at the two synchronism frequencies with main lobe bandwidth
of $\Delta f_{1,2} \approx \frac{1}{\tau _{sp_{1,2}}}$, where
$\tau_{sp_{1,2}} \approx
\left| \frac{L_w}{v_{z_0}} - \frac{L_w}{v_{g_{1,2}}} \right|$
is the slippage time.
The corresponding temporal wave-packet
(shown in Fig.\ \ref{two_solutions_energy}b)
consist of two ``slow'' and ``fast'' pulses with durations equal to the
slippage times modulating carriers at their respective synchronism frequencies.
The spectral bandwidth in the case of grazing
shown in Fig.\ \ref{grazing_energy}a,
is determined by dispersive effects of the waveguide taking into account by the
simulation.
The corresponding temporal wavepacket is shown in Fig.\ \ref{grazing_energy}b.
\\

\noindent
{\bf Acknowledgments}

The research of the second author (Yu.~L.) was supported in part by
the Center of Scientific Absorption of the Ministry of Absorption,
State of Israel.


\begin{table}[b]
\caption{The operational parameters of millimeter wave free-electron maser.}
\label{table}
\begin{center}
\begin{tabular}{ll} 
\multicolumn{2}{l}{\underline{Accelerator}} \\
  Electron beam energy:  & $E_k$=1$\div$3~MeV \\
  Electron beam current: & $I_0$=1~A \\
  Pulse duration:        & $T$=1~pS \\[2mm]
\multicolumn{2}{l}{\underline{Wiggler}} \\
  Magnetic induction:    & $B_w$=2000~G \\
  Period:                & $\lambda_w$=4.444~cm \\
  Number of periods:     & $N_w$=20  \\[2mm]
\multicolumn{2}{l}{\underline{Waveguide}}  \\
  Rectangular waveguide: &  1.01~cm $\times$ 0.9005~cm \\
  Mode:                  &  $TE_{01}$   \\  
\end{tabular}
\end{center}
\end{table}

\clearpage

\clearpage

\end{document}